\documentclass[twocolumn,amsmath,superscriptaddress,amsfonts,prb]{revtex4}
\usepackage{graphicx}
\pdfoutput=1
\begin{document}
\title{Field induced nucleation in nano-structures}
\author{V. G. Karpov}\email{victor.karpov@utoledo.edu}\affiliation{Department of Physics and Astronomy, University of Toledo, Toledo, OH 43606, USA}
\author{D. Niraula}\email{dipesh.niraula@rockets.utodeo.edu}

\begin{abstract}
We predict the probability of field induced nucleation (FIN) of conductive filaments across the nano-thin dielectric layers in memory and switching devices. The novelty of our analysis is that it deals with a dielectric layer of thickness below the critical nucleation length. We show how the latter constraint can make FIN a truly threshold phenomenon possible only for voltage (not the field) exceeding a certain critical value that does not depend on the dielectric thickness. Our analysis predicts the possibility of threshold switching without memory under certain thickness dependent voltages. In parallel, the thermal runaway mechanism of electronic switching is described analytically leading to results consistent with the earlier published numerical modeling. Our predictions offer the possibility of experimental verifications deciding between FIN and thermal runaway switching.
\end{abstract}
\maketitle

\section{Introduction}
The presence of conductive filaments (CFs) is critically important for functionality of phase change memory (PCM), \cite{PCM} resistive random access memory (RRAM) \cite{RRAM}, and threshold switches (TS) \cite{TS}. Also, CFs are responsible for the breakdown phenomena in gate dielectrics.

While the important role of CFs is commonly recognized, their underlying physics is not sufficiently understood. The two types of mechanisms of CF formation have been proposed in the literature: (1) the direct electric field induced nucleation (FIN) of a new phase in the form of conductive filamentary pathway characterized by its critical length and nucleation barrier, \cite{karpov2008,karpov2017} and (2) the electronic filament precursor raising the local temperature enough to trigger a phase transformation or remain as such for the case of TS. \cite{gallo2016}

It is worth recalling here that CFs in the above mentioned devices are initially created by the electro-forming process that requires a certain forming voltage $\sim 1-4$ V. A formed structure is then modified (using additional electric stimuli) in such a way as to introduce a relatively narrow insulating gap across the filament. Repeatedly closing and opening that gap makes the device switching between two distinct states underlying its functionality. As the order of magnitude estimates, we mention the original CF length of $H\sim 10-20$ nm, and the insulating gap width $h_{\rm gap}\sim 1-3$ nm. \cite{pickett2009,cagli2017,long2012} The voltages required to create and bridge the insulating gaps in CFs are typically several times lower than the forming voltages. \cite{long2012,larentis2012,valov2013}

\begin{figure}[b!]
\includegraphics[width=0.43\textwidth]{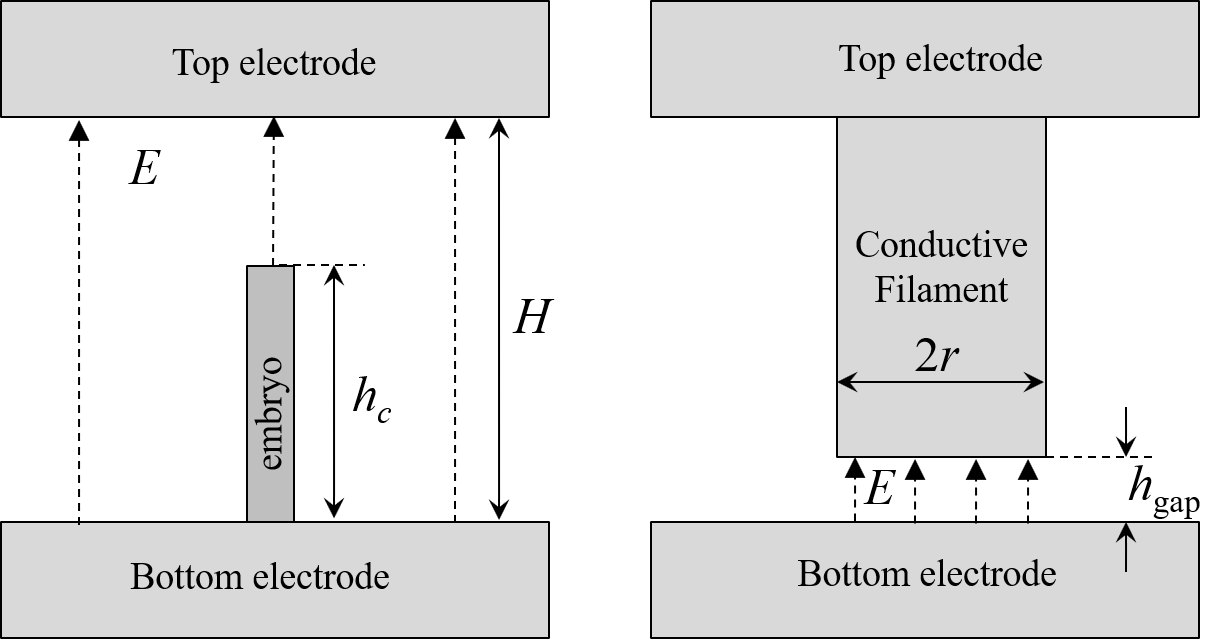}
    \caption{Left: critical length ($h_c$) conductive embryo formed in a uniform electric field within a gap of height $H$ significantly exceeding $h_c$. That condition corresponding to that of the standard FIN theory can relate to a pristine (before `forming') device structure. Right: a conductive filament with an insulating gap of thickness $h_{\rm gap}$ in a `formed' device. A small conductive embryo will nucleate inside the gap of $h_{\rm gap}<h_{c}$, i. e. beyond the limits of applicability of the standard FIN. Dashed arrows represent the electric field.}\label{fig:setup}
    \end{figure}

Here we consider a constrained FIN where the volume available for nucleation is limited. That condition can be important for the modern nanometer devices where conductive bridges form through  narrow, $h_{\rm gap}\sim 1-3$ nm gaps,  while the critical nucleation length $h_c\gg h_{\rm gap}$, say, $h_c\sim 10$ nm, \cite{karpov2008,karpov2017} as illustrated in Fig. \ref{fig:setup}. Also, it can be relevant for the electroforming processes if the dielectric layer thickness is small enough, $\lesssim 10$ nm.

The paper is organized as follows. In Sec. \ref{sec:FIN} we apply the standard field induced nucleation theory to the narrow ($h_c\gg h_{\rm gap}$) gap case. Sec. \ref{sec:nFIN} will introduce a different FIN scenario driven by the free energy originating from the effective capacitor formed by the CF tip and the opposite electrode. Sec. \ref{sec:disc} discusses the interplay between the two FIN scenarios. The thermal runaway electronic mechanism of switching is analyzed in Sec. \ref{sec:electr}. Sec. \ref{sec:concl} contains final conclusions.

\section{The standard FIN scenario}\label{sec:FIN}
We recall that FIN is a process where  a small metallic needle-shaped embryo nucleates in a non-conductive host due to the polarization energy gain in a strong external electric field $E$. Similar to other nucleation processes, FIN is characterized by the critical nucleation length $h_c$ and nucleation barrier $W$ representing respectively the length above which the embryo grows spontaneously, and its corresponding energy. Along the standard lines, the latter quantities are defined without any spacial constraints.

As demonstrated earlier, \cite{nardone2009} nucleation in TS and non-volatile memory is described similarly; for the sake of specificity, here, we consider the case of TS. Assuming a high aspect ratio nucleus, $h\gg 2r$  its free energy in a uniform field of strength $E$ is presented as,
\begin{equation}\label{eq:FIN}
F(h,r)=2\pi rh\gamma -\alpha h^3\varepsilon E^2.
\end{equation}
where $\gamma$ is the surface tension. The second term in Eq. (\ref{eq:FIN}) represents the electrostatic energy gain and is written in a truncated form where $\alpha$ stands for a combination of numerical and logarithmic multipliers. For example, $\alpha =\{12\ln[(2h/r)-1]\}^{-1}$ for the ellipsoidal CF, and $\alpha =\{\ln[(2h/r)-7]\}^{-1}$ for the cylindrical CF, assuming $h\gg r$. \cite{landau1984, karpov2008} In what follows we neglect the logarithmic dependence of $\alpha$ treating it as  a constant smaller than one, say, $\alpha \sim 0.1$.

The free energy of Eq. (\ref{eq:FIN}) has a stationary point, which is a maximum and thus unrelated to the transition barrier between the initial state ($r=h=0$) and the new phase state where $F$ decreases with $h$ and $r$. It was shown that such a barrier is provided by the absolute minimum of $F(h,r)$ located at the boundary region of the minimum allowed value of $r$ (in sub-nanometer range) denoted here as $r_{\rm min}$. The latter is determined by the conditions of electric and mechanical integrity for CF. \cite{karpov2008}
%

\begin{figure}[t!]
\includegraphics[width=0.4\textwidth]{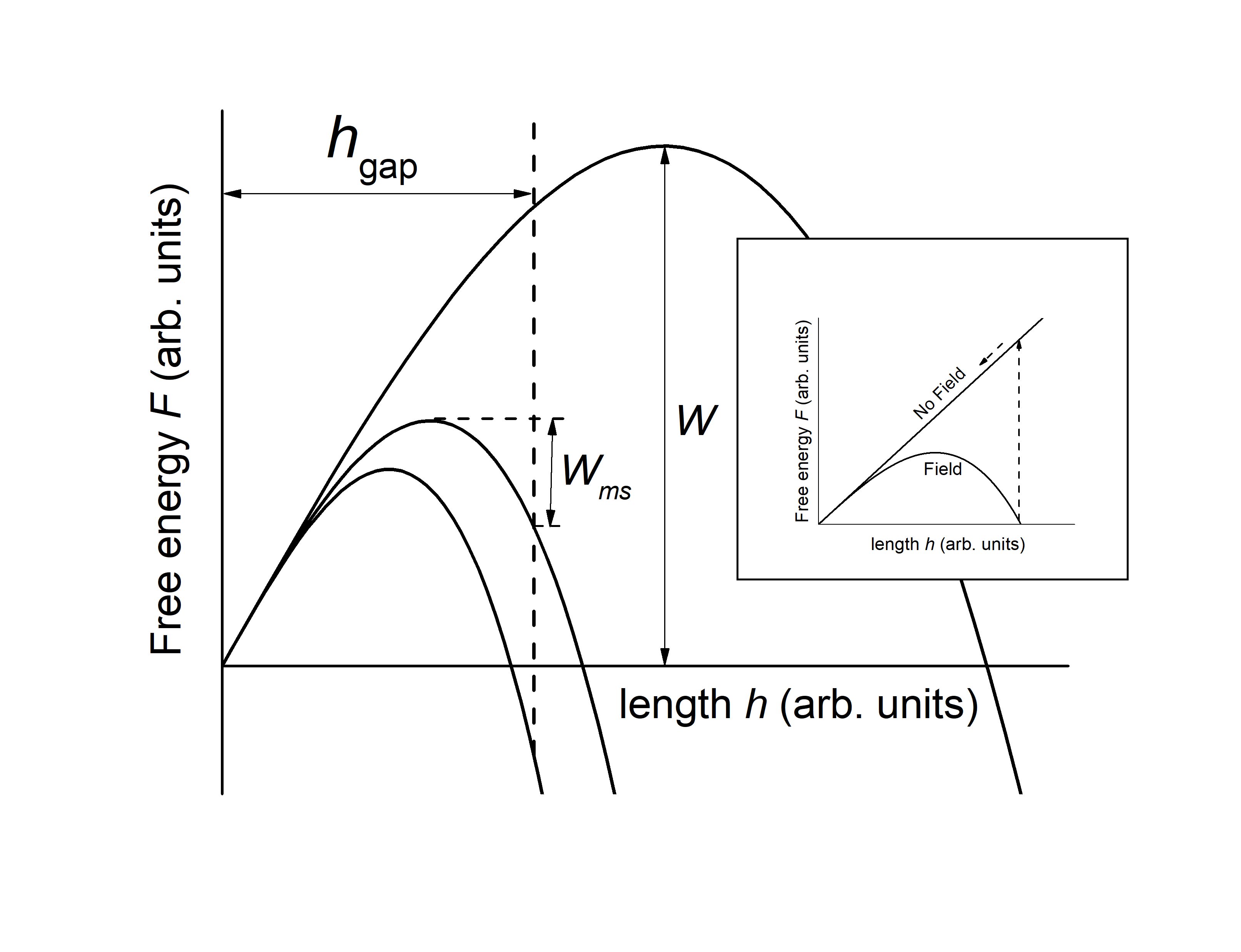}
    \caption{The free energies of a conductive embryo vs. its height in the electric fields of three different strengths, corresponding (from right to left) to the unstable, metastable, and thermodynamically stable embryos. The initial state of the system corresponds to $h\approx 0$ (to the accuracy of thermal energy fluctuation $\sim kT$). $W$ is the nucleation barrier height, $W_{ms}$ is the barrier determining the lifetime of a metastable embryo. The inset diagram shows the transition to unstable state upon the field removal.\label{fig:CF}}
    \end{figure}

For $r=r_{\rm min}$, the free energy is a minimum at
\begin{equation}\label{eq:hc} h_c =\sqrt{\frac{2\pi r_{\rm min}\gamma}{3\varepsilon \alpha}}\frac{1}{E}\end{equation}
corresponding to the barrier height
\begin{equation}\label{eq:W}
W(E)=F[h_c(E),r_{\rm min}]=\frac{2^{5/2}(\pi r_{\rm min}\gamma )^{3/2}}{3^{3/2}(\alpha \varepsilon )^{1/2}E}\end{equation}
as illustrated in Fig. \ref{fig:CF}.

For the future references, it is convenient to represent the nucleation barrier through the electric potential $U=EH$,
\begin{equation}\label{eq:W1}W(U)=W_0\frac{U_0}{U}\end{equation}
with
\begin{equation}\label{eq:defs}W_0=2\pi\gamma r_{\rm min}H,\quad {\rm and}\quad  U_0=\sqrt{\frac{8\pi r_{\rm min}\gamma}{27\alpha\varepsilon}}.\end{equation}
Because the nucleation time $\tau =\tau _0\exp(W/kT)$ where $\tau _0={\rm const}$, it follows that FIN takes place under any voltage $U$ over the time that exponentially decreases with $U$ as
\begin{equation}\label{eq:tau}\tau \propto \exp[(W_0/kT)(U_0/U)].\end{equation} This type of dependence was experimentally observed. \cite{bernard2010,sharma2015,yoo2017}

Considering FIN in a narrow gap, the weak field region, $E<E_{ms}$ [explicitly given in Eq. \ref{eq:ms})] can be determined by the condition $h_c>h_{\rm gap}$. In that region, the free energy within the gap is increasing with $h$ making embryos unstable. On the other hand, under higher fields, $E>E_s$ [also, explicitly given in Eq. \ref{eq:ms})], the barrier is at shorter distance than $h_c$, and yet, the energy $F(h_{\rm gap}, r_{\rm min})$ is positive. Therefore,
for the field strengths in the interval,
\begin{equation}\label{eq:ms}
E_{ms}\equiv   \sqrt{\frac{2\pi r_{\rm min}\gamma}{3\varepsilon \alpha h_{\rm gap}^2}}<E<\sqrt{\frac{2\pi r_{\rm min}\gamma}{\varepsilon \alpha h_{\rm gap}^2}}\equiv E_s, \end{equation}
a metastable embryo within a gap ($h_c<h_{\rm gap}$) can be formed. Shown in Fig. \ref{fig:CF} the barrier determining that embryo lifetime is given by
\begin{equation}\label{eq:Wms} W_{ms}=W(E)-F(h_{\rm gap})\end{equation}
with $W(E)$ and $F(h)$ from Eqs. (\ref{eq:W}) and (\ref{eq:FIN}). The metastable embryo lifetime is given by
\begin{equation}\label{eq:taums}\tau _{ms} =\tau _0\exp(W_{ms}/kT)\end{equation} with $W_{ms}=0$ when $E=E_{ms}$.

The upper characteristic field $E_s$ in Eq. (\ref{eq:ms}) is determined by the condition $F(h_{\rm gap})=0$, i.e. the embryo remains stable as long as the field $E$ is applied. When the field is removed, the second term in Eq. (\ref{eq:FIN}) disappears, and the embryo free energy linear in $h$ makes it decay as illustrated in Fig. \ref{fig:CF} relevant for TS functionality.

It follows from the above that while the nucleation of unconstrained CF is possible for the field of any strength (with exponentially field dependent nucleation times), the confined CFs require the field strength $E>E_{ms}$ and the corresponding voltages $U>U_{ms}=E_{ms}h_{\rm gap}$. Similarly, stable CFs can nucleate when $U>U_{s}=E_{s}h_{\rm gap}$. It follows from Eq. (\ref{eq:ms}) that both $U_s$ and $U_{ms}$ are independent of $h_{\rm gap}$ and $U_s=\sqrt{3}U_{ms}$. Therefore, as opposed to the case of unconstrained CF, the confinement seems to make nucleation a truly threshold voltage phenomenon.

Interestingly, the latter expressions for $U_s$ and $U_{ms}$ predict numerical values consistent with the data in the order of magnitude, \cite{fantini2012,chen2012,wouters2012,wei2015,sawa2005,beck2000}
\begin{equation}\label{eq:est}U_s=\sqrt{\frac{2\pi\gamma r_{\rm min}}{\varepsilon\alpha}}=\left(\frac{3^{3/2}}{2}\right)U_0\sim  0.3-1 \quad {\rm V}.\end{equation}
Here and in what follows we use the numerical values listed in Table \ref{tab:param}.

Also, we note that the transition from the regime of constant field to constant voltage with dielectric thinning below 10 nm was observed in PCM devices, \cite{yu2008} although its original explanation was different. It was demonstrated that in 30 nm PCM structures switching is dominated by the field strength rather than voltage. \cite{wimmer2014}

\begin{table}[h]
\caption{Some parameters related to FIN}
\begin{tabular}{|l|c|c|c|c|c|c|}
   \hline
  Parameter & $H$, nm\footnotemark[1]& $r_{\rm min}$, nm\footnotemark[2] & $\varepsilon$\footnotemark[3] &$\gamma$, dyn/cm\footnotemark[4] &$\alpha$\footnotemark[5] & $\kappa$, cm$^2$/s\footnotemark[6] \\  \hline
  Value & 3 & 0.3 &25 & 10-100 & 0.1 & 0.1\\ \hline
 \end{tabular}

\footnotetext[1]{Following published estimates. \cite{pickett2009,cagli2017,long2012}}
\footnotetext[2]{We use $r_{\rm min}$ discussed in the early work on FIN. \cite{karpov2008}}
\footnotetext[3]{We use the dielectric permittivity of HfO$_2$.}
\footnotetext[4]{Because the values of interfacial energies in materials undergoing FIN are not available, we use the ballpark of typical values for a variety of other systems. \cite{jeurgens2009,israelachvili1992} }
\footnotetext[5]{See the discussion after Eq. (\ref{eq:FIN})}
\footnotetext[6]{Thermal diffusivity estimated or HfO$_2$ based RRAM devices.\cite{karpov2017}}
\label{tab:param}\end{table}

\section{Another scenario of FIN}\label{sec:nFIN}
The preceding section analysis tacitly assumed the polarization energy of a metal needle remaining cubic in length regardless of its closeness to opposite electrode. That assumption remains valid through almost the entire gap, since the electric field is strongly different from the uniform only in a small region $\sim r^2/h\ll h_{\rm gap}$ around the tip. \cite{karpov2014s}

However, there is another gap related effect significantly contributing to the free energy: capacitive interaction between the CF tip and its opposite electrode. That interaction can be thought of as due to a flat plate capacitor formed by the tip's end face and the opposite electrode, leading to the energy contribution,
\begin{equation}\label{eq:FC}F_C=-\frac{CU^2}{2} \approx -\frac{\varepsilon r^2U^2}{8h_{\rm gap}}\end{equation}
where $C$ is the above mentioned capacitance approximated with that of the flat plate capacitor, $U$ is the voltage between the electrodes. Note that assuming other geometrical shapes of the tip lead to slightly different results, \cite{hudlet1998,ruda2003} where, for example, $r$ is replaced with the electrode size in Eq. (\ref{eq:FC}) for the case of hyperboloid of revolution. \cite{ruda2003}

A comment is in order to explain the negative sign of $F_C$ in Eq. (\ref{eq:FC}). In the process of changing the gap capacitance $\Delta C$ with its width $h_{\rm gap}$, we assume that the system remains at constant voltage, so the capacitor energy changes by $U^2\Delta C/2$. However, simultaneously the charge $\Delta Q=U\Delta C$ passes through the power source, which requires the energy $-U\Delta Q= -U^2\Delta C$ making the total energy change equal $-U^2\Delta C/2$.
\begin{figure}[t!]
\includegraphics[width=0.45\textwidth]{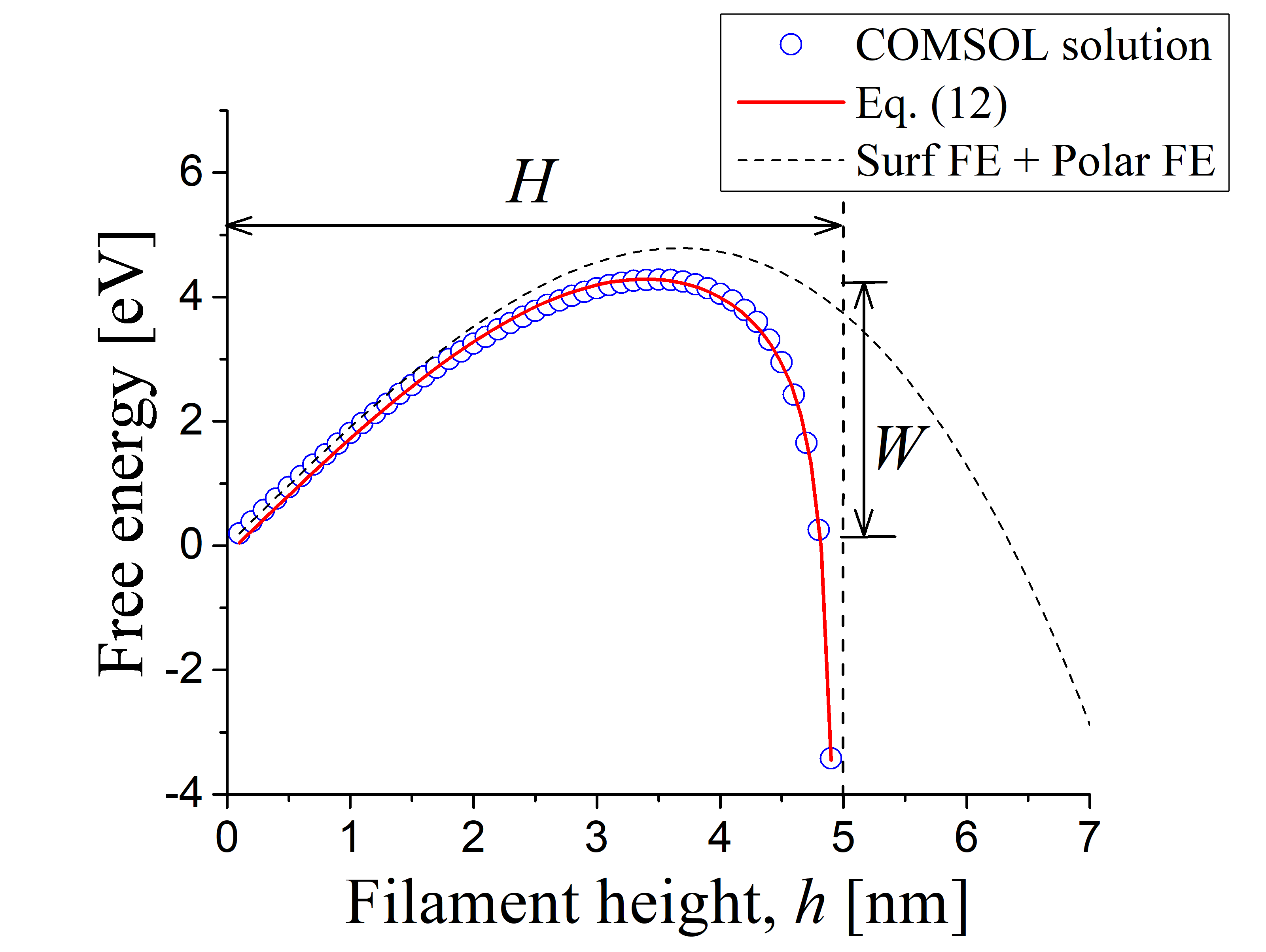}
    \caption{The free energy of CF in a narrow gap between two electrodes. Symbols represent the results of the COMSOL numerical modeling for 0.5 nm radius CF between the two coaxial circular metal electrodes of 10 nm radius each. The solid line is a fit by Eq. (\ref{eq:nFIN}). The dashed line is the free energy corresponding to the same surface tension and polarization, but without the capacitive interaction.\label{fig:CFgap}}
    \end{figure}

Based on Eqs. (\ref{eq:FIN}) and (\ref{eq:FC}) and assuming the TS case, the free energy of CF can be approximated as
\begin{equation}\label{eq:nFIN}
F(h,r)=2\pi\gamma rh -\alpha h^3\varepsilon (U/H)^2-\frac{\varepsilon r^2U^2}{8(H-h)}.
\end{equation}
A major feature added here is a sharp energy decrease for relatively small gap widths $h_{\rm gap}\ll H$ due to interaction between the CF tip and opposite electrode. We have verified the above heuristic free energy form of Eq. (\ref{eq:nFIN}) with COMSOL modeling for a variety of CF dimensions; one example is presented in Fig. \ref{fig:CFgap} for a cylinder shaped CF of radius $r=0.5$ nm between two circular electrodes of radii 10 nm each. Overall, the analytical form of free energy in Eq. (\ref{eq:nFIN}) provides a description within $\sim 10$ \% of relative error.

The free energy of Eq. (\ref{eq:nFIN}) has a stationary point at certain $h$ and $r$ that can be determined analytically. Here we skip the corresponding cumbersome equations noting that, for that point, $D\equiv\partial ^2F/\partial h^2 \partial ^2F/\partial r^2-(\partial ^2F/\partial h\partial r)^2>0$ and $\partial ^2F/\partial h^2<0$, $\partial ^2F/\partial r^2<0$, which identifies it as a maximum.

The case of a maximum stationary point in free energy landscape is similar to that of the standard field induced transformation scenario described in the preceding section: the phase transformation barrier is determined by the minimum acceptable radius $r_{\rm min}$. Its related gap width is given by $dF(h,r_{\rm min})/dh=0$ with $F$ from Eq. (\ref{eq:nFIN}). Because the polarization term [second on the right hand side of Eq. (\ref{eq:nFIN})] is relatively unimportant for $H-h\ll H$, we get,
\begin{equation}\label{eq:gap}H-h=H\frac{U}{2U_{\gamma}}\quad {\rm where}\quad U_{\gamma}\equiv\sqrt{\frac{4\pi\gamma H^2}{\varepsilon r_{\rm min}}}.\end{equation}
The corresponding barrier height is given by,
\begin{equation}\label{eq:barrier}W=W_0 (1-U/U_{\gamma}),\end{equation}
leading to the nucleation time exponentially dependent on bias. We conclude that the above introduced capacitive interaction acts as a sort of clutch decreasing CF energy in a narrow interval of its heights close to the opposite electrode. The corresponding nucleation barrier decreases with voltage linearly unlike the standard FIN dependence in Eq. (\ref{eq:tau}).

Note that the parameters from Table \ref{tab:param}, yield a numerical estimate of the characteristic voltage $U_\gamma\sim 3$ V.  The corresponding gap width from Eq.(\ref{eq:gap}) is typically in sub-nanometer range. For small enough voltages, Eq.(\ref{eq:gap}) predicts $h_{\rm gap}\lesssim 1$ {\AA}, which raises a question on applicability of this theory , since that narrow gaps would allow  significant electric currents, energy dissipation, and electric shorting.

To understand the role of the latter effects, we note the very short thermalization time $\tau _T\sim h_{\rm gap}^2/\kappa \lesssim 0.1$ ps where $\kappa$ is the thermal diffusivity ensuring local quasi-equilibrium. The dissipation will increase the quasi-equilibrium temperature, thus accelerating the nucleation. On the other hand, the concomitant electric shorting, would decrease the electrostatic energy of the entire device towards its final value corresponding to the fully formed CF. That decrease aggravates the free energy falloff again helping nucleation.

The quantitative description of the above mentioned transient temperature increase and shorting effect fall  beyond the present scope. Therefore, we should admit a degree of uncertainty making the proposed FIN scenario questionable for small enough voltages. If that scenario does not work, the structures with nano-gaps can work only as TS and not memory when their applied biases are below $U_{ms}$. The viability of $\sim 1$ {\AA} gaps between the nucleated CF and opposite electrode may depend on the morphology of materials involved.

\section{Interplay of two FIN scenarios}\label{sec:disc}
\begin{figure}[t!]
\includegraphics[width=0.25\textwidth]{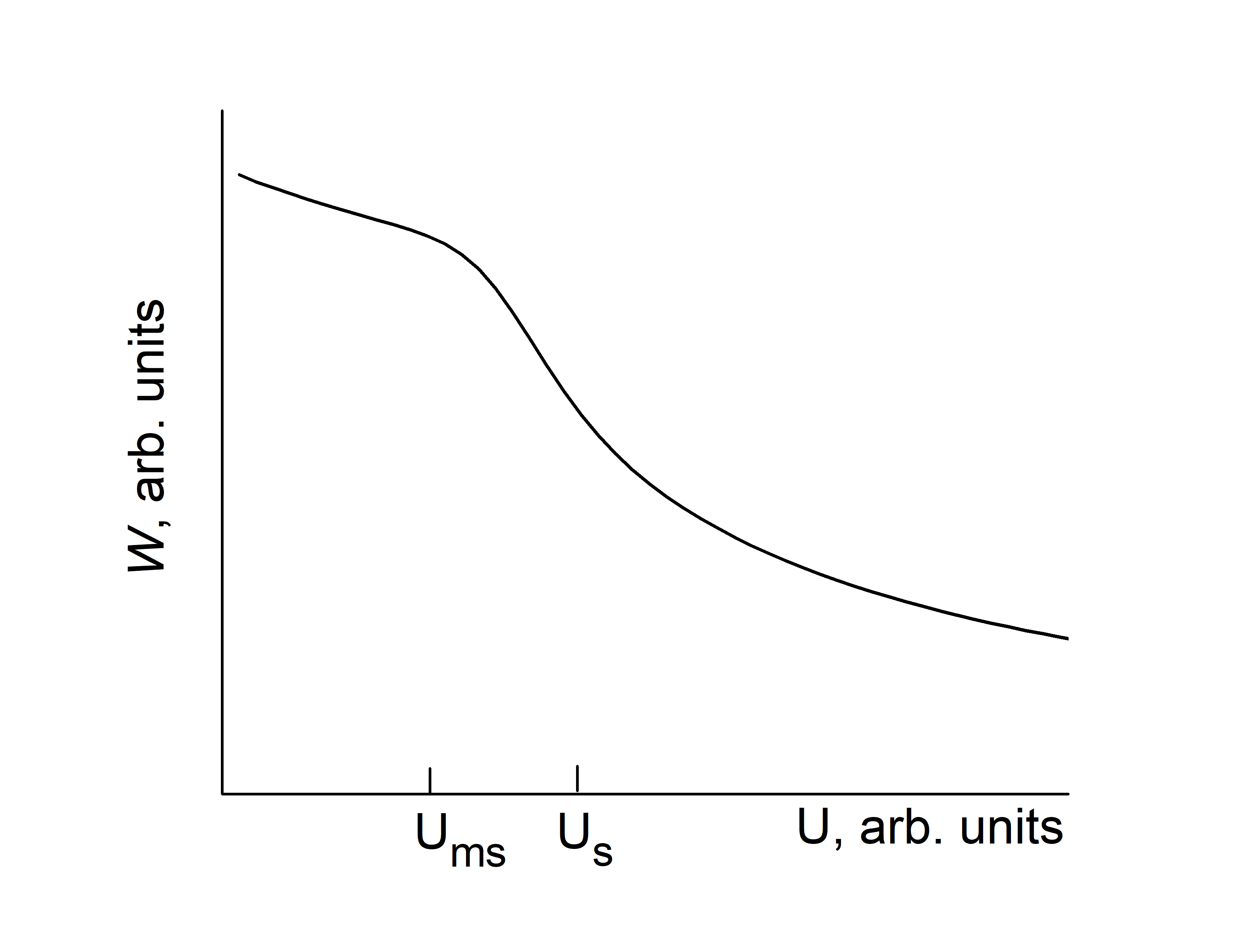}\includegraphics[width=0.25\textwidth]{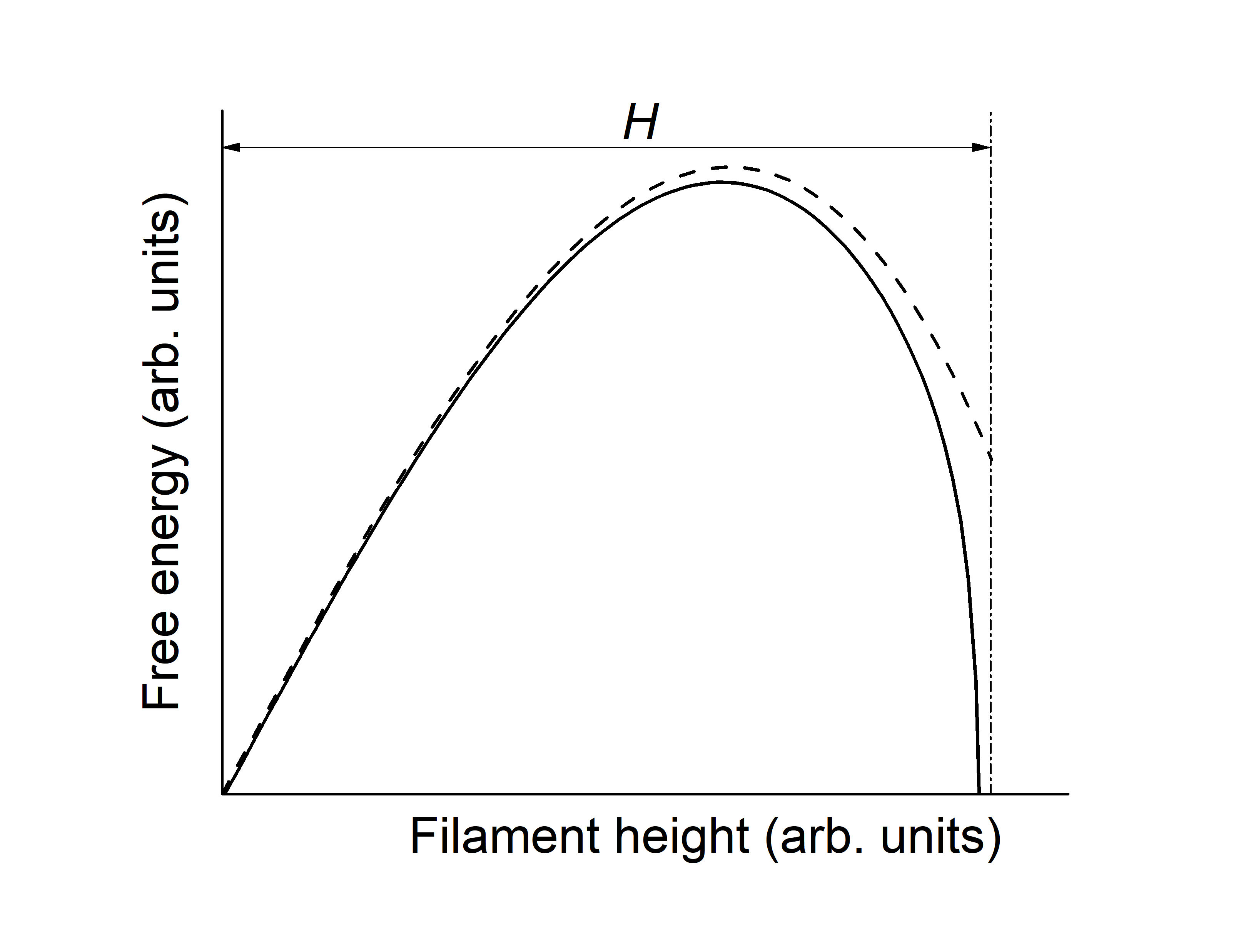}
    \caption{{\it Left}: A sketch of the FIN barrier voltage dependence, linear in low voltage region and hyperbolic for higher voltages. {\it Right}: The filament free energies with (solid line) and without (dashed line) the tip electrode interaction [the last term in Eq. (\ref{eq:nFIN})] taken into account.\label{fig:Nucbar}}
    \end{figure}

The interplay between the standard and here developed FIN models is described by two relations,
\begin{equation}\label{eq:interp1}
\frac{U_{\gamma}}{U_s}=\sqrt{2\alpha}\frac{H}{r_{\rm min}}\end{equation}
and
\begin{equation}\label{eq:interp2}\frac{W_{\rm new}}{W_{\rm stand}}=\left(\frac{3^{3/2}}{2}\right)\frac{U(1-U/U_{\gamma})}{U_s}\end{equation}
where the nucleation barriers $W_{\rm stand}$ and $W_{\rm new}$ are given respectively by Eqs. (\ref{eq:W1}) and (\ref{eq:barrier}).
Comparing the barrier shapes in Figs. \ref{fig:CFgap} and \ref{fig:CF} we then conclude that the standard FIN scenario dominates (i. e. $W_{\rm stand}< W_{\rm new}$) when $U> U_c\equiv 2/(3)^{3/2}U_s$. That is the same range of voltages where the standard FIN theory predicts the voltage dependence of $\tau$ in Eq. (\ref{eq:tau}). We recall, in addition, that the transition between the two FIN scenarios has a threshold nature at voltage $U_{ms}=E_{ms}h_{\rm gap}$ [see Eq. (\ref{eq:ms})]. That is slightly above $U_c$, since, based on their definitions, $U_c=(2/3)U_{ms}$. Such a behavior is illustrated in Fig. \ref{fig:Nucbar} (Left).

It should be noted in addition that the criterion of nucleation in the standard FIN scenario depends on the post-nucleation CF evolution and conditions determining the desired lifetime of the initially metastable CF created under voltage above $U_{ms}$. In general, the sufficient voltage is determined by the condition that the nucleated CF survives the desired time as specified in Eq. (\ref{eq:taums}). However, the results presented in Sec. \ref{sec:nFIN} show that  the tip-electrode interaction can transform the metastable CF into a stable one as illustrated in Fig. \ref{fig:Nucbar} (Right).

The above consideration predicts that, given all other factors, the switching voltage should decrease when the gap $H$ decreases. Indeed, using $\tau =\tau _0\exp(W/kT)$ along with Eq. (\ref{eq:barrier}), the voltage capable of triggering nucleation over time $\tau$ becomes,
\begin{equation}\label{eq:utau}
U(\tau )=\left[H-\frac{kT}{2\pi\gamma r_{\rm min}}\ln\left(\frac{\tau}{\tau _0}\right)\right]\sqrt{\frac{4\pi\gamma}{\varepsilon r_{\rm min}}},\end{equation}
It increases with $H$ for any given exposure time. The latter prediction is consistent with the data \cite{long2012,larentis2012,valov2013} showing how the switching voltage in a formed structure is by a factor 2-4 lower than the forming voltage.

\begin{figure}[b!]
\includegraphics[width=0.37\textwidth]{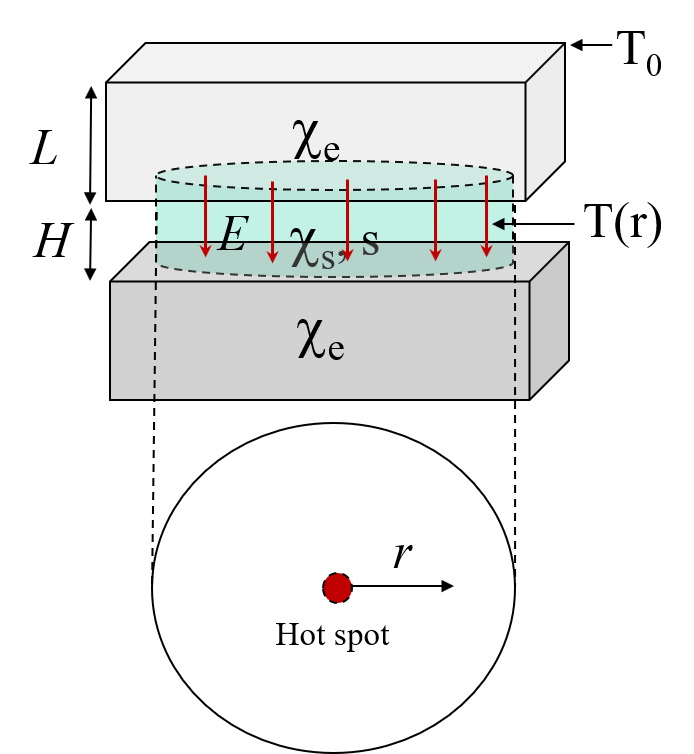}
    \caption{A sketch of a semiconductor layer between two metal electrodes of thickness $L$ each. The bottom view shows a hot spot caused by the runaway instability.\label{fig:thermal}}
    \end{figure}

\section{Electronic switching}\label{sec:electr}
A number of recent publications  \cite{li2016,goodwill2017,li2017,funck2017,funck2016,legallo2016} theoretically studied purely switching (without any structural transformations) that might pertain to threshold switches operating as selector devices in modern solid state memory arrays. Their underlying thermal runaway scenario utilizes a non-linear current voltage characteristic where the conductivity is thermally and field activated,
\begin{equation}\label{eq:cond}\sigma =\sigma _0\exp(-w/kT) \quad {\rm with}\quad w=w_0-\delta w(E).\end{equation}
Here $\delta w(E)$ is the field induced decrease in the activation energy of conductivity. A local lateral variation of current will then generate excessive local heat and temperature additionally increasing the current density at that location, etc.; hence, thermal electron instability evolving into a narrow filament carrying high electric current. The formation of such a filament is identified as switching.

Because the underlying modeling \cite{li2016,goodwill2017,li2017,funck2017,funck2016,legallo2016} remains numerical, here, we give a simple analytical treatment allowing to examine the corresponding parameter ranges. We start with the standard 3D heat transfer equation $c\partial T/\partial t=\chi\nabla ^2T+\sigma E^2=0$ where the thermal capacitance ($c$), thermal conductivity ($\chi$), and electric conductivity taking their respective values in the semiconductor and metal electrode materials forming a structure with axial symmetry shown in Fig. \ref{fig:thermal}.

Given the ambient temperature $T_0$ at the external surfaces and the Neumann boundary conditions at the electrode-semiconductor interfaces, it was shown \cite{niraula2017} that averaging the semiconductor temperature along the transversal direction (between the electrodes) reduces the heat transport equation to the form
\begin{equation}\label{eq:ht}\frac{c_s}{\chi _s}\frac{\partial T}{\partial t}=\nabla ^2T+\beta ^2(T_0-T)+\frac{\sigma E^2}{\chi _s},\quad \beta ^2 \equiv \frac{2\chi _e}{\chi _sHL}.\end{equation}
Here $T$ depends on the radial coordinate $r$, indexes $s$ and $e$ denote respectively the semiconductor and electrode materials. $\beta$ is a reciprocal thermal length characterizing the decay of radial nonuniformities. Note that the second and third terms on the right-hand-side represents respectively heat dissipation and evolution.

Eq. (\ref{eq:ht}) along with Eq. (\ref{eq:cond}) allows the standard linear stability analysis constituting a natural approach to studying run-away phenomena. Introducing the average lateral temperature $\overline{T}$ and substituting
\begin{equation}\label{eq:Fourier}\delta T\equiv T-\overline{T}=\sum _qT_q\exp(i\omega _qt-iqr)\quad (\ll \overline{T}),\end{equation} yields in the linear approximation,
\begin{equation}\label{eq:Tav}(\overline{T}-T_0)\frac{\beta ^2\chi _s}{\sigma _0E^2}=\exp\left(-\frac{w}{k\overline{T}}\right)\end{equation} and
\begin{equation}\label{eq:disp}\frac{c_s}{\chi _s}i\omega _q=-q^2-\beta ^2+\frac{\sigma _0 E^2}{\chi _s}\frac{w}{k\overline{T}^2}\exp\left(-\frac{w}{k\overline{T}}\right).\end{equation}
The instability takes place for positive real values of $i\omega$. For the modes with $q$ below a certain wave number $q_0$, the latter criterion reduces to the form,
\begin{equation}\label{eq:inst}\lambda\equiv\frac{\overline{T}-T_0}{\overline{T}}\frac{w}{k\overline{T}}\frac{\beta ^2}{\beta ^2+q_0^2}\geq \lambda _{cr}=1.\end{equation}
While derived differently, the inequality in Eq. (\ref{eq:inst}) is similar to the classical criterion \cite{frank1969,landau1987} $\lambda\geq \lambda _{cr}\approx 0.88$ for 1D thermal instability with $q_0=0$ and $\beta =H/2$.

\begin{figure}[t!]
\includegraphics[width=0.37\textwidth]{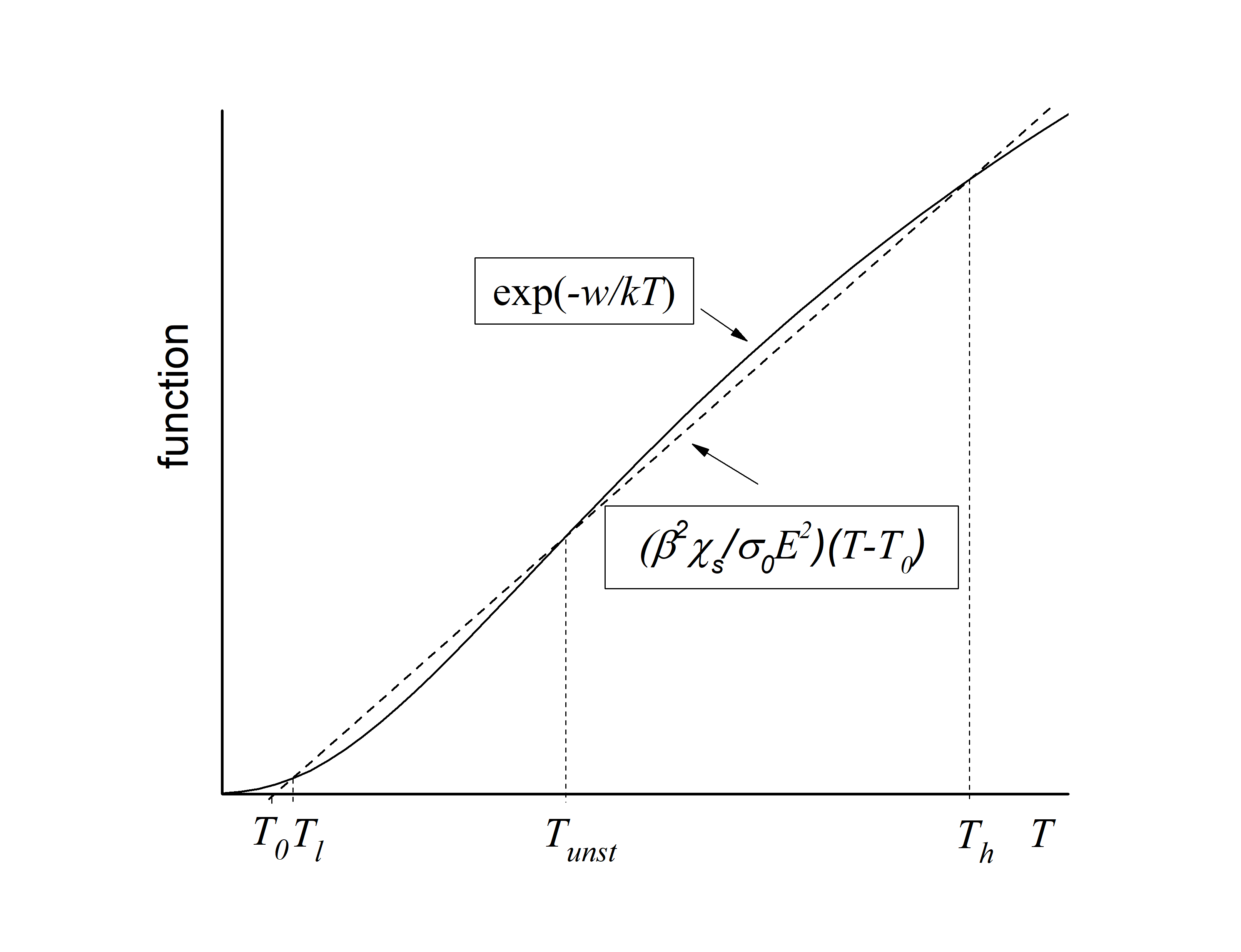}
    \caption{A sketch of the graphical solution of Eq. (\ref{eq:Tav}) where the straight line and the curve represent its left- and right-hand sides.\label{fig:graphsol}}
    \end{figure}

A graphical representation in Fig. \ref{fig:graphsol} shows that when the field is low enough, Eq. (\ref{eq:Tav}) has three solutions: low ($T_l$) corresponding to the uniform current flow, high ($T_h$) for the hot spot, and intermediate ($T_{unst}$), which is unstable as seen from the relationship between the heat evolution and dissipation [cf. the note after Eq. (\ref{eq:ht})].  That conclusion agrees with the general phenomenological analysis of thermal instabilities. \cite{subashiev1987} Furthermore, the condition for critical field (straight line tangential to the exponential curve in Fig. \ref{fig:graphsol}) self-consistently coincides with the criterion in Eq. (\ref{eq:inst}) when $q=0$.

The above simplistic analysis provides approximate numerical results given the same parameters as in the published numerical modeling. \cite{li2016,goodwill2017,li2017,funck2017,funck2016,legallo2016} For example,  assuming values \cite{funck2016} $E=10^8$ V/m, $\sigma _0=10 ^7$ S/m, $\chi _s=0.12$ Wm$^{-1}$K$^{-1}$, $H=10$ nm, $L=30$ nm, $\chi _e/\chi _s = 170$, $R=60$ nm, yields $\beta ^2\approx 10^{18}$ m$^{-2}$, and $w/k\overline{T}=\ln\{\sigma _0E^2/[\beta ^2\chi _s (\overline{T}-T_0)]\}\approx 7$. Here we have rather arbitrarily approximated $(\overline{T}-T_0)=T_0$ under the logarithm, which has no significant effect when it appears with a multiplier that is by many orders of magnitude bigger than $(\overline{T}-T_0)$.  Substituting into Eq. (\ref{eq:inst}) gives $7(\overline{T}-T_0)/\overline{T}=1$, i. e. the switching temperature $\overline{T}\approx 340$ K, in fare agreement with numerical simulations. \cite{funck2016}

Furthermore, the estimated $\overline{T}\approx 340$ K and $w/k\overline{T}=7$ yield $w\approx 0.21$ eV, which must be interpreted as the barrier for the above used $E=10^8$ V/m. Any additional assumptions about  that barrier field dependence are not necessary as long as $w$ remains an adjustable parameter, which was the case for all the preceding numerical modeling \cite{li2016,goodwill2017,li2017,funck2017,funck2016,legallo2016} carried out under the assumptions of $\delta w(E)$ following the Pool-Frenkel law. We would like to emphasize here that the thermal runaway switching will take place for any model of thermally activated conductivity with or without field or voltage dependent barrier. In particular, fitting the data with Pool-Frenkel law based modeling does not appear indicative of that law.  Of course, the switching field will depend on that barrier assumptions when the value of zero field barrier $w_0$ is postulated.

Some additional observations are as follows.\\ (i) Using the available data, Eq. (\ref{eq:inst}) shows that switching takes place when the barrier is still significant, $w/kT\gg 1$. With that in mind, one can approximate the required barrier suppression vs. gap width, $\delta w(E)={\rm const}+kT\ln H$, which dependence may be too weak to resolve experimentally. \\
(ii) Unlike FIN, the runaway model per se does not predict any delay time between voltage pulse and switching. The observed delay was attributed in that model to the thermalization time \cite{li2016} or left without explicit interpretation. \cite{legallo2016} As long as the measured delay time is exponential in voltage, the only relevant mechanism is the conductivity activation where $\tau\propto \exp [-\delta w(E)/T]$; hence, the prediction of delay time vs. voltage being reciprocal of the voltage dependent conductivity, is open for experimental verifications. Note that FIN predictions for $\tau (U)$ are  quite different as specified in Eqs. (\ref{eq:tau}) and (\ref{eq:barrier}).\\
(iii) The above analysis will describe the finite area effects when we set $q_0=2\pi /R$ where $R$ is the device radius taking into account, along the general lines, that the strongly oscillating terms with $q>2\pi /R$ in the expansion of Eq. (\ref{eq:Fourier}) are immaterial. With that in mind, Eqs. (\ref{eq:Tav}) and (\ref{eq:inst}) predict that switching requires temperature increase $(\overline{T}-T_0)$ that are by the factor of $1+4\pi ^2(\beta ^2R^2)^{-1}$ greater than that for infinitely large devices. Since $(\overline{T}-T_0)\propto E^2$, we conclude that the switching fields and potentials scale as $\sqrt{1+4\pi ^2(\beta ^2R^2)^{-1}}$ with device size, which is at least qualitatively consistent with the results of numerical modeling. \cite{li2017} That scaling becomes practically important for nano-sized devices. No such scaling is predicted by FIN where switching field remains independent of device size.

The above (ii) and (iii) can be used to experimentally decide between the switching mechanisms of FIN and thermal runaway. A general concern about the latter arises for the case of submicron and especially nanometer thick amorphous structures with transversal conduction varying by many orders of magnitude between different spots, \cite{raikh1987} which strongly affect the thermal runaway mechanism. \cite{karpov2012}

\section{Conclusions}\label{sec:concl}
We have shown that, \\
(1) FIN in narrow (shorter than the critical nucleation length) gaps between two metal electrodes exhibits new trends governed by certain characteristic voltages in the parctically interesting range.\\
(2) There exists the characteristic voltage $U_{ms}$ independent of the gap width, below which the nucleation barrier decreases with voltage linearly, while it is reciprocal of voltage above $U_{ms}$, and the transition is of threshold nature independent of the gap width.\\
(3) For some materials, nucleation of CF can result in functionality of only TS without memory.\\
(4) Our analytical consideration of thermal runaway switching provides results consistent with the earlier published numerical modeling and leads to several predictions offering experimental verifications deciding between the mechanisms of FIN and thermal runaway.\\

This work was supported in part by the Semiconductor Research Corporation (SRC) under Contract No. 2016-LM-2654. We are grateful to I. V. Karpov and R. Kotlyar for useful discussions.

\end{document}